\documentclass[pdflatex,sn-mathphys-num]{sn-jnl}

\usepackage{graphicx}%
\usepackage{multirow}%
\usepackage{amsmath,amssymb,amsfonts}%
\usepackage{amsthm}%
\usepackage{mathrsfs}%
\usepackage[title]{appendix}%
\usepackage{xcolor}%
\usepackage{textcomp}%
\usepackage{manyfoot}%
\usepackage{booktabs}%
\usepackage{algorithm}%
\usepackage{algorithmicx}%
\usepackage{algpseudocode}%
\usepackage{listings}%
\usepackage{xspace}
\usepackage{natbib}
\usepackage{courier}
\usepackage{lineno}

\newcommand{\model}{BrainWave\xspace}
\newcommand{\labram}{LaBraM\xspace}
\newcommand{\brainbert}{BrainBERT\xspace}
\newcommand{\moment}{MOMENT\xspace}
\newcommand{\vpara}[1]{\vspace{0.01in}\noindent\textbf{#1 }}

\theoremstyle{thmstyleone}%
\theoremstyle{thmstyletwo}%

\theoremstyle{thmstylethree}%

\begin{document}

\title[]{\model: A Brain Signal Foundation Model for Clinical Applications}


\author[1]{\fnm{Zhizhang} \sur{Yuan}}\email{zhizhangyuan@zju.edu.cn}

\author[1]{\fnm{Fanqi} \sur{Shen}}\email{fanqishen@zju.edu.cn}

\author[2]{\fnm{Meng} \sur{Li}}\email{li.meng@mail.sim.ac.cn}

\author[4,5]{\fnm{Yuguo} \sur{Yu}}\email{yuyuguo@fudan.edu.cn}

\author[1]{\fnm{Fei} \sur{Wu}}\email{wufei@zju.edu.cn}

\author[6]{\fnm{Chenhao} \sur{Tan}}\email{chenhao@uchicago.edu}

\author*[1]{\fnm{Yang} \sur{Yang}}\email{yangya@zju.edu.cn}

\affil[1]{\orgdiv{Computer Science and Technology}, \orgname{Zhejiang University}, \orgaddress{\city{Hangzhou}, \state{Zhejiang}, \country{China}}}

\affil[2]{\orgdiv{Shanghai Institute of Microsystem and Information Technology}, \orgname{Chinese Academy of Sciences}, \orgaddress{\city{Shanghai}, \country{China}}}


\affil[4]{\orgdiv{Research Institute of Intelligent and Complex Systems, State Key Laboratory of Medical Neurobiology and MOE Frontiers Center for Brain Science, and Institute of Science and Technology for Brain-Inspired Intelligence}, \orgname{Fudan University}, \orgaddress{\city{Shanghai}, \country{China}}}

\affil[5]{\orgdiv{Shanghai Artificial Intelligence Laboratory}, \orgname{Fudan University}, \orgaddress{\city{Shanghai}, \country{China}}}

\affil[6]{\orgname{University of Chicago}, \orgaddress{\city{Chicago}, \state{Illinois}, \country{USA}}}




\abstract{
Neural electrical activity is fundamental to brain function, underlying a range of cognitive and behavioral processes, including movement, perception, decision-making, and consciousness.
Abnormal patterns of neural signaling often indicate the presence of underlying brain diseases.
The variability among individuals, the diverse array of clinical symptoms from various brain disorders, and the limited availability of diagnostic classifications, have posed significant barriers to formulating reliable model of neural signals for diverse application contexts.
Here, we present \model, the first foundation model for both invasive and non-invasive neural recordings, pretrained on more than 40,000 hours of electrical brain recordings (13.79 TB of data) from approximately 16,000 individuals.
Our analysis show that \model outperforms all other competing models and consistently achieves state-of-the-art performance in the diagnosis and identification of neurological disorders.
We also demonstrate robust capabilities of \model in enabling zero-shot transfer learning across varying recording conditions and brain diseases, as well as few-shot classification without fine-tuning, suggesting that \model learns highly generalizable representations of neural signals.
We hence believe that open-sourcing \model will facilitate a wide range of clinical applications in medicine,
paving the way for AI-driven approaches to investigate brain disorders and advance neuroscience research.
}




\maketitle

\section{Introduction}

Electrical brain recordings, capturing the intricate patterns of the brain's electrical activities, are essential in advancing our understanding of brain across scientific domains~\cite{Barborica2023Studying,jiang2020spatially,engel2005invasive,pesaran2018investigating,urai2022large,khodagholy2015neurogrid,horejs2024long}. 
In particular, they can be used to identify medical conditions and diagnose neurological disorders, and are thus essential for addressing major global health challenges in developing nations~\cite{Shih2012brain, Feigin2020global}.
Scalp electroencephalography (EEG) and intracranial electroencephalography (iEEG) are two primary methods to conduct these types of recordings.
EEG is non-invasive and economical viable, and has thus been used in diverse applications~\cite{Soufineyestani2020Electroencephalography,Varbu2022Past,Jadhav2022Clinical,Silva2024EEG,Amer2023EEG}. 
In contrast, iEEG offers high signal fidelity and spatial resolution~\cite{Yamada2024scalable}, but the invasive nature limits its applicability to only the most severe patient cases and restricted scenarios.
Due to the distinct features of EEG and iEEG data, such as different acquisition rates and notable variations in channel numbers~\cite{dasgupta2022previous,Parvizi2018Promises,lachaux2003intracranial}, studies have so far investigated them separately. 
We hypothesize that combining EEG and iEEG data can offer information that are not only rich in detail but also highly generalizable across diverse neural electrical activities,
and develop a pioneering foundational model, \model, 
for both EEG and iEEG data.
\model learns robust representations that achieve the state-of-art performance in a wide range of tasks, demonstrating the synergy of EEG and iEEG data for the first time.

\model 
overcomes a suite of inherent challenges in conventional supervised artificial intelligence (AI) models used for the analysis of brain signals.
First, \model leverages self-supervised training, circumventing the need for large-scale, high-quality manual labeling.
In clinical applications, the process of brain data annotation is labor-intensive and reliant on specialized expertise~\cite{bertalan2019data, pascual2019self,Zhao2023Classification}, exemplified by the need for multi-day monitoring for epilepsy patients~\cite{friedman2009how} and the clinical experts' capacity to annotate only tens of seconds of data in a single work period. 
Second, \model provides much-need generalization at two levels that were not possible in supervised training, which requires an understanding of fundamental and general patterns in brain signals.
Individual variability in brain neural activities, a consequence of each person's distinct brain structure and functional behaviors~\cite{Brown2017Individual}, leads to markedly different brain recording patterns~\cite{yuan2023ppi}. 
This diversity hinders the generalizability of supervised AI models, as they often struggle to extend the insights gained from a subset of patients to a broader population, due to significant differences across individuals, as well as variability that changes with different behavioral states. 
Moreover, there are numerous types of brain-related diseases with various underlying mechanisms ~\cite{clemente2023mitochondria,mcewen2015mechanisms,delgado2017epigenetic,gaiteri2014beyond}, and even a single disease may present with multiple subtypes~\cite{yang2021temporal}.
Supervised AI models are also task-specific and fail to address a diverse range of tasks using brain signals~\cite{guo2021detecting,WANG2022Sexplainable,Chen2022brainnet,yuan2023ppi,BAGHERZADEH2022Detection,SAHU2023automated,Miltiadous2023Novel,Vicchietti2023Computational,Sun2024Multi}, indicating that they do not actually ``understand'' brain signals.

\begin{figure}
\begin{center}

\includegraphics[width=\linewidth]{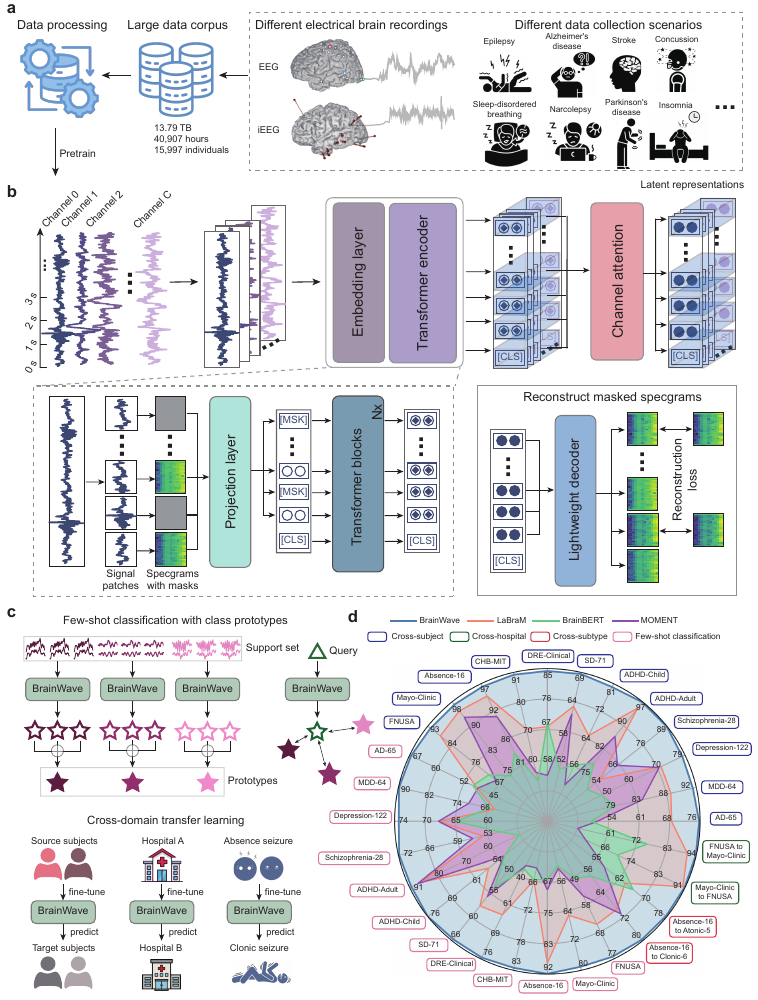}

\caption{
    \label{fig1}
    \textbf{Overview of \model}.
    \textbf{a,} Data curation for pretraining \model. The pretraining corpus contains both invasive and non-invasive brain recordings collected from diverse healthcare scenarios.
    \textbf{b,} The pretraining pipeline of \model. \model is pretrained on more than 3 billion signal patches using a masked modeling strategy.
    \textbf{c,} The evaluation tasks consist of few-shot classification and cross-domain evaluation. We conduct few-shot classification with a prototypical network in which the we directly compare the representations of the queries with class prototypes. We perform three different levels of cross-domain analysis: cross-subject, cross-hospital, and cross-subtype. 
    \textbf{d,} The overall results of \model compared to other pretrained models. \model outperforms other models across all the 28 experiments, with significant improvement ($p < 0.001$) in 24 of them.
}

\end{center}
\end{figure}

While prior work has attempted to build foundation models for brain signals~\cite{Zhou2023foundation, Chen2024Towards, Xu2024whole, Pai2024Foundation,Zhang2024generalist,Hao2024Large}, \model offered unique technical contributions
by building the largest dataset of electrical brain recordings and developing novel techniques to integrate EEG and iEEG data for the first time. 
As illustrated in Figure~\ref{fig1}a, we collected a total of 13.79 TB of combined EEG and iEEG data over a duration of 40,907 hours, which serves as the foundation for pretraining \model.
The data were obtained from 15,997 individuals, including both healthy individuals and those with various brain disorders, spanning an age range from infancy ($<1$ year) to over 90 years old.
In the pretraining stage, we adopt a masked modeling strategy that reconstructs the time-frequency representations of the masked patches (Fig.~\ref{fig1}b).
Unlike previous works that uniformly resampled data to a common sampling rate~\cite{jiang2024large,zhang2023brant,wang2023brainbert}, the designed embedding layer of \model can adapt to different temporal resolutions, which enhanced the data scalability and provided the model with stronger generalization capabilities.
We also employed a channel count-agnostic approach to capture the inter-channel relationships.
Empowered by growing datasets and advances in model design, \model exhibited highly robust pretrained representations and excellent transfer learning capabilities (Fig.~\ref{fig1}c).

\begin{figure}
\begin{center}

\includegraphics[width=\linewidth]{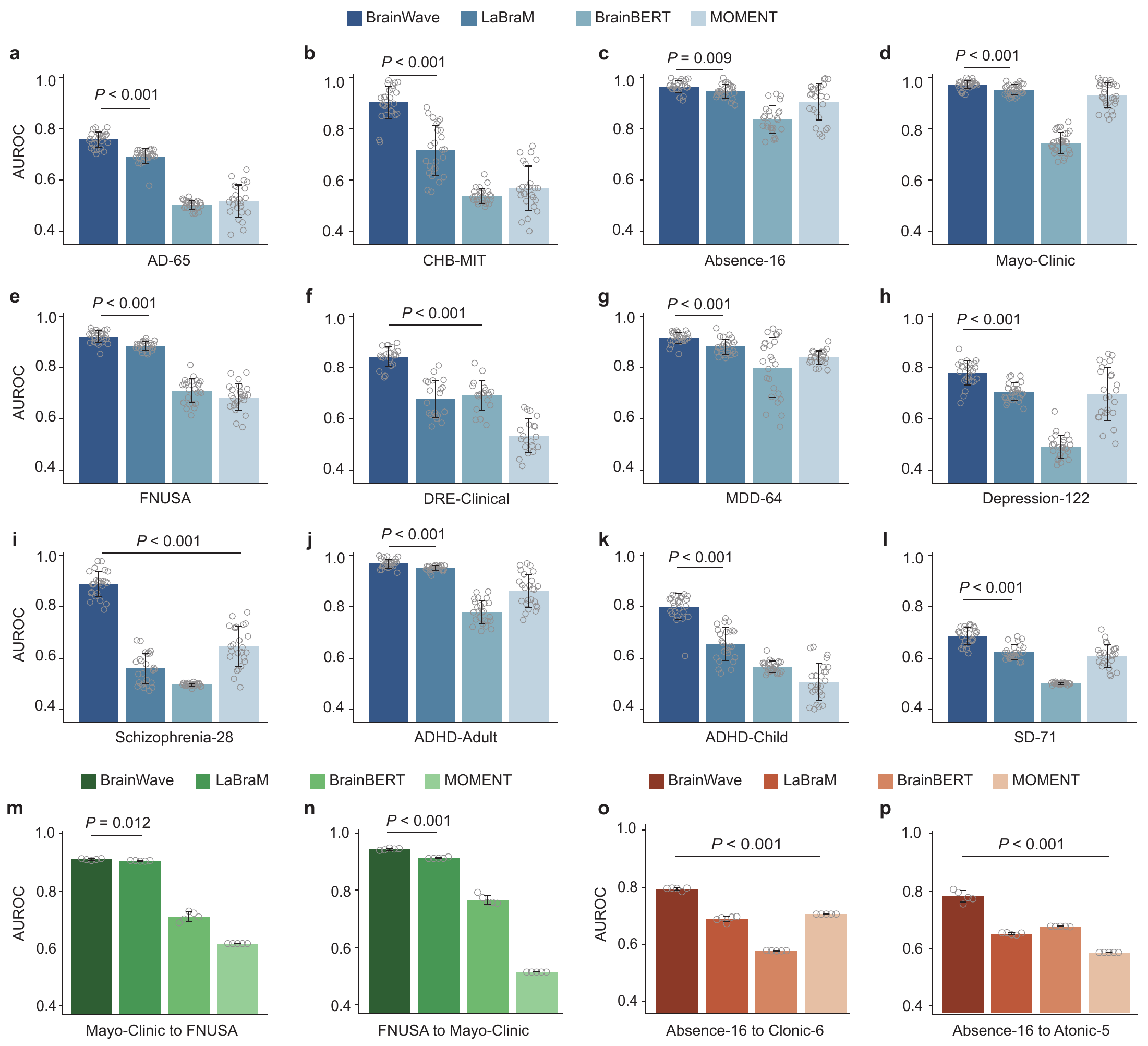}

\caption{
    \label{fig2}
    \textbf{Performance of cross-domain evaluation.}
    \textbf{a-l,} Bar plots comparing the AUROC scores of \model and competing models on cross-subject tasks.
    Each experiment is conducted with $n$-fold cross validation ($n$ is the number of subject groups), where we repeat five runs for each fold. 
    \textbf{m,n,} Bar plots comparing the AUROC scores of \model and competing models on cross-hospital tasks.
    \textbf{o,p,} Bar plots comparing the AUROC scores of \model and competing models on cross-subtype tasks.
    \textbf{m-p,} The source dataset is served as the training set and the target dataset is served as the evaluation set. In each experiment, we repeat five runs.
    \textbf{a-p,} Data are mean $\pm$ SD. The listed $p$ value indicates the significance for \model outperforming the best comparison model, with the two-sided $t$-test.
}

\end{center}
\end{figure}

We systematically evaluated \model and observed its exceptional performance across different settings, including cross-subject (Fig.~\ref{fig2}a and b), cross-hospital (Fig.~\ref{fig2}c) and cross-subtype (Fig.~\ref{fig2}d) tasks, which attests to its strong generalization capabilities. Additionally, its proficiency in few-shot classification (Fig.~\ref{fig3}) highlights its adaptability in learning from limited data. 
We also demonstrated the effectiveness of joint pretraining by leveraging both invasive and non-invasive neural data (Fig.~\ref{fig4}a-c) and found that it achieved superior performance by learning more generalizable representations (Fig.~\ref{fig4}e) and richer semantic information (Supplementary Tables 20 and 21).
To comprehensively assess the capabilities of \model as a foundation model for electrical brain recordings in healthcare scenarios, we constructed a benchmark comprising 15 different datasets.
\model is compared against the previous state-of-the-art foundation models that are publicly available, including \labram~\cite{jiang2024large}, \brainbert~\cite{wang2023brainbert} and \moment~\cite{goswami2024moment}.
Figure~\ref{fig1}d summarizes the overall results of \model compared with other methods, in which \model attains consistently state-of-the-art performance on all the 28 experiments, with significant improvement ($p < 0.001$) over the second-best method in 24 experiments, showing the versatility of \model in a wide array of tasks for brain recordings in healthcare.
To investigate the effectiveness of joint pretraining with both invasive and non-invasive neural data, we separately pretrained two model variants using EEG and iEEG data under the condition of maintaining other pretraining settings constant, and compared them across all experimental settings (Fig.~\ref{fig4}a-d). 
The results show that \model outperforms other variants in diverse tasks and experimental setups, marking it as the first to successfully implement joint pretraining and validate its effectiveness.  
Overall, we demonstrated the potential of \model to assist clinical diagnostics and decision support with strong scalability.

\model not only exhibits potential for broad applications but also offers new insights for the development of foundation models based on brain signals.
We will release \model as an off-the-shelf and publicly available model, which will serve as a basis for others in their own tasks, facilitating diverse clinical applications and research for brain recordings.

\section{Results}

\subsection{Cross-domain Disease Diagnosis and Detection}
In real-world clinical applications, where the data of individuals to be diagnosed are usually unavailable during training, it is desirable for a model to be capable of making accurate predictions on unseen subjects.
Thus, evaluating in a cross-subject setting is essential to accurately reflect a model's performance and practical value in clinical scenarios.
For each dataset, we split the subjects into $n$ non-overlapping groups and employed $n$-fold cross validation to evaluate all models, ensuring that all subjects are included in the testing process.
In each fold, we randomly selected a subject group from the training set for validation and repeated five runs.
\model was evaluated against competing methods on a total of 12 datasets, consisting of 10 publicly available datasets (Supplementary Tables 7-16) and 2 private datasets (Supplementary Tables 3 and 4).
Model performance was reported using the area under the receiver operating curve (AUROC) and balanced accuracy (BACC).
We calculated $p$ values with the two-sided $t-$test between \model and the most competitive comparison model for each task to check for significance.

Across all experiments, \model consistently outperformed other methods with an average relative improvement of 12.61\% in AUROC (Area Under the Receiver Operating Characteristic curve) and 16.44\% in BACC (Balanced Accuracy) over each second-best performing model (Fig.~\ref{fig2}a-l and Extended Data Fig.~\ref{cross_sub_bacc}).
In schizophrenia diagnosis (Fig.~\ref{fig2}i; dataset Schizophrenia-28~\cite{Schizophrenia-28}), \model achieved an improvement of 37.44\% and 41.59\% compared to the best comparison model in terms of AUROC and BACC. 
In seizure detection (Fig.~\ref{fig2}b; dataset CHB-MIT~\cite{CHB-MIT}), \model demonstrated a 26.18\% boost relative to the second best method.
The strong performance of \model on both ADHD-Adult~\cite{ADHD-adult} (Fig.~\ref{fig2}j) and ADHD-Child~\cite{ADHD-child} (Fig.~\ref{fig2}k) indicated its robustness across age groups, owing to the broad age distribution in our pretraining corpus.
In conclusion, \model significantly surpassed other models ($p<0.001$) on 11 out of the 12 datasets, demonstrating its superiority in disease diagnosis and detection.

The promising results of \model on cross-subject evaluation motivated us to further explore its transfer capability across more divergent distributions.
Thus, we attempted two more challenging experimental setups. 
Under these settings, we fine-tuned on one dataset and directly apply the model to another. 
These datasets are not only collected from different individuals but also from different hospitals and collection devices, or from patients with different disease subtypes.
Unlike the traditional paradigm where fine-tuning is dataset-specific, these settings allow the model to be seamlessly deployed across different datasets and even different but related tasks.
The cross-hospital evaluation involved mutual transfer between two datasets, Mayo-Clinic and FNUSA~\cite{nejedly2020multicenter}, both collected from patients with drug resistant epilepsy but from different hospitals.
The cross-subtype evaluation included three datasets, namely Absence-16, Clonic-6, and Atonic-5, collected from patients with different subtypes of seizures (absence seizure, clonic seizure, and atonic seizure), and we performed zero-shot transfer from Absence-16 to Clonic-6 and Atonic-5.
\model showed promising results and achieved the best performance in all cross evaluations (Fig.~\ref{fig2}m-p and Extended Data Fig.~\ref{cross_bacc}).
For instance, \model achieved an impressive AUROC of 93.82\% in the zero-shot transfer from FNUSA to Mayo-Clinic (Fig.~\ref{fig2}n).
When transferring across different seizure subtypes (Fig.~\ref{fig2}o,p), \model exhibited average improvements of 13.90\% and 13.49\% in terms of AUROC and BACC, respectively, compared to the second-best model.

In summary, our experimental results indicated that \model holds the potential to reduce labeling and training costs under similar scenarios, demonstrating promising applicability in real-world clinical tasks.

\begin{figure}
\begin{center}

\includegraphics[width=\linewidth]{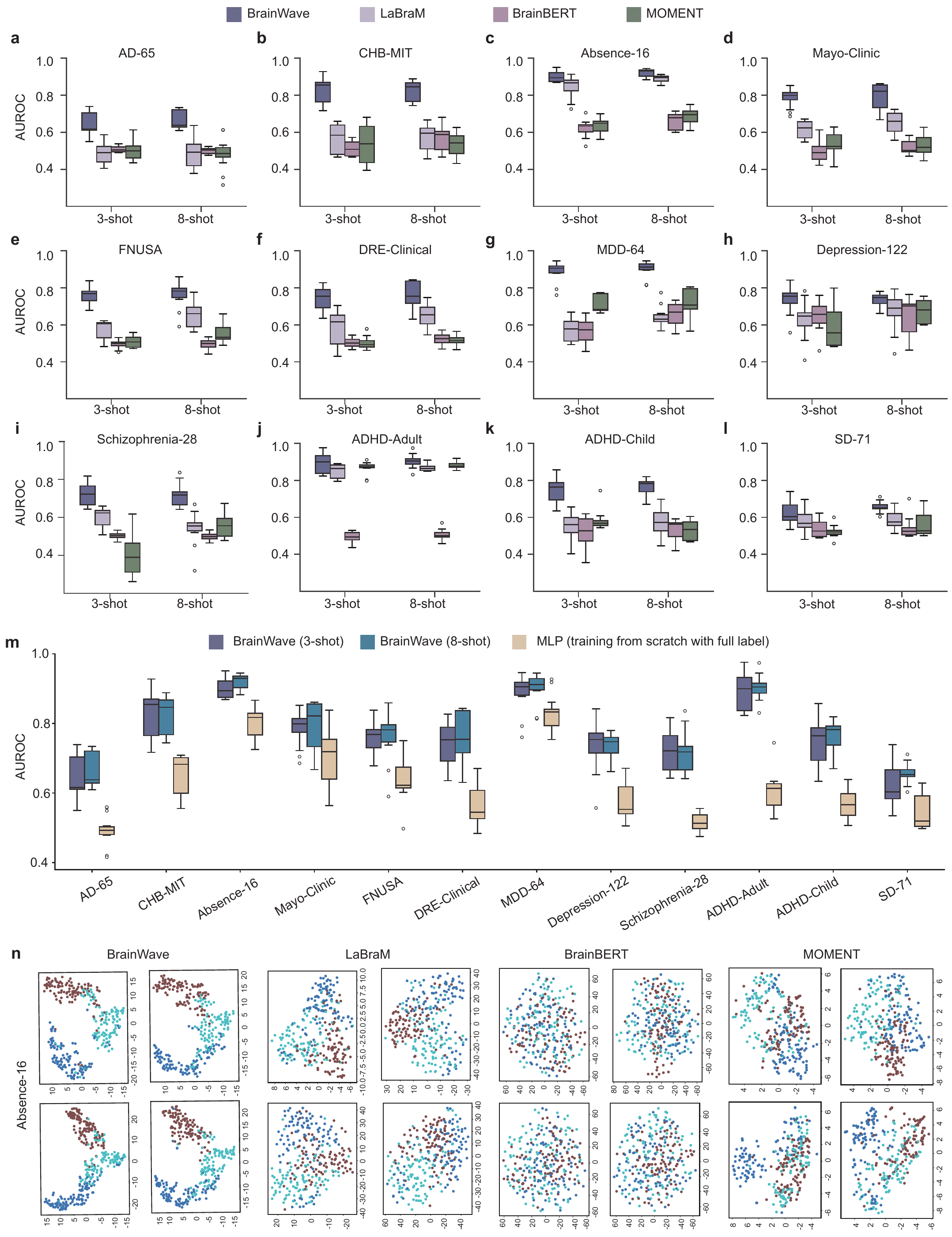}

\caption{
    \label{fig3}
    \textbf{Performance and analysis of few-shot classification.}
    \textbf{a-l,} Box plots comparing the AUROC scores of \model and competing models on few-shot classification. We conduct $n$-fold cross validation for each experiment and repeat five runs per fold. We perform 3-shot and 8-shot classification for each task.
    \textbf{m,} Box plots comparing the BACC scores of \model on few-shot classification and end-to-end trained MLP with full-label supervision.
    \textbf{n,} t-SNE (t-distributed Stochastic Neighbor Embedding) plots of the pretrained representations on Absence-16 generated from \model and other pretrained encoders. Each model contains four subplots, with each subplot generated by randomly sampling a portion of the original dataset.
}

\end{center}
\end{figure}
\subsection{Few-shot Classification}
In clinical practice, limited availability of labeled data sometimes poses challenges for fine-tuning models, which highlights the critical importance of learning sufficiently robust representations.
Consequently, we performed few-shot classification, which is an evaluation scheme that studies the generalization capabilities of models on new tasks given a very limited number of labeled examples.
We adopted a direct comparison strategy for classification by comparing the representations of the queries with prototype of each category (Fig.~\ref{fig1}c). 
The process solely involved obtaining representations from the pretrained models and computing class prototypes, without any parameter updates or introduction of new parameters.
The few-shot experiments were still conducted under the cross-subject setting and we employed $n$-fold cross validation, where in each fold we randomly chose labeled examples from the training set as the support set.
We established two sizes for the support set, with 3 and 8 labeled examples per class (3-shot and 8-shot), respectively.
Given that the performance can fluctuate depending on the support set, we repeated experiments over five runs in each fold.
For comparison, we also trained an MLP (multilayer perceptron) from scratch with full-label supervision for each experiment.

We conducted few-shot experiments (Fig.~\ref{fig3}a-l and Extended Data Fig.~\ref{few_shot_bacc}) on all datasets used in the cross-subject evaluation and found that \model still maintained solid performance by achieving an average improvement of 21.21\% in terms of AUROC compared to the second-best model.
For instance, on Absence-16 (Fig.~\ref{fig3}c) and ADHD-Adult (Fig.~\ref{fig3}j), \model achieved an AUROC over 90\% (91.93\% on Absence-16, 90.39\% on ADHD-Adult) in 8-shot classification.
On MDD-64~\cite{mumtaz2016mdd} (Fig.~\ref{fig3}g), the performance of 8-shot learning even nearly matched that of full-label supervised fine-tuning (89.83\% versus 91.50\%).
\model not only outperformed other models by a large margin but also demonstrated greater robustness to the selection of the support set. 
Specifically, we calculated the average standard deviation across all few-shot experiments and found that the fluctuations in \model's performance are, on average, smaller than those of the second-best performing models (5.74\% versus 6.06\%).
The poor performance of \brainbert in few-shot classification may be due to its limitation in supporting only fixed-length inputs, which requires additional operations to accommodate varying input lengths, highlighting the importance of supporting variable-length inputs for a wide array of tasks.

When compared to the end-to-end trained MLP (Fig.~\ref{fig3}m), \model still achieved an average improvement of 25.79\% and 26.60\% in terms of AUROC and BACC, respectively.
Surprisingly, we also observed that when comparing the 8-shot performance of \model with the full-label fine-tuning (i.e., with thousands or even tens of thousands of labeled examples) performance of other pretrained models, our model still outperforms on the majority of datasets (Extended Data Figs.~\ref{8shot_finetune_auc} and~\ref{8shot_finetune_bacc}).
Specifically, \model surpassed \labram, \brainbert and \moment on 7, 12 and 11 out of 12 datasets, respectively.
To better illustrate the results, we visualized the representations of the pretrained models, with different colored points representing different categories (Fig.~\ref{fig3}n) and Extended Data Fig.~\ref{tsne}). 
Even without fine-tuning, the representations generated by \model are sufficiently discriminative and enable its outstanding performance in few-shot classification.
Overall, our extensive evaluation of few-shot classification demonstrated the immense potential of \model as a foundational model for electrical brain recordings that offers robust and off-the-shelf representations for various clinical applications.


\begin{figure}
\begin{center}

\includegraphics[width=\linewidth]{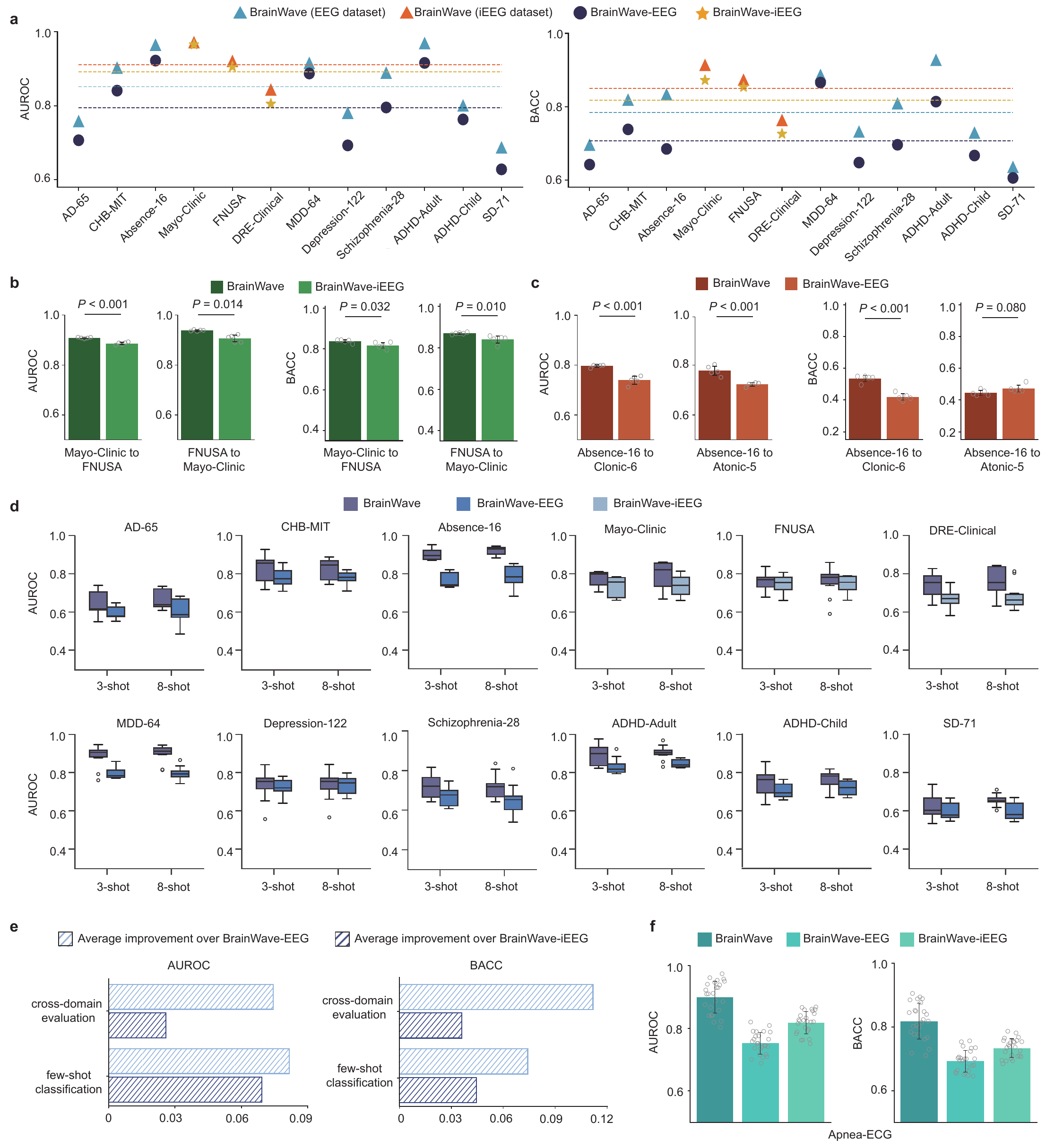}

\caption{
    \label{fig4}
    \textbf{Analysis of joint pretraining. }
    \textbf{a,} Scatter plots comparing the AUROC and BACC scores of \model, \model-EEG and \model-iEEG on cross-subject tasks.
    \textbf{b,} Bar plots comparing the AUROC and BACC scores of \model and \model-iEEG on cross-hospital tasks.
    \textbf{c,} Bar plots comparing the AUROC and BACC scores of \model and \model-EEG on cross-subtype tasks.
    \textbf{b,c,} The source dataset is served as the training set and the target dataset is served as the evaluation set. In each experiment, we repeat five runs. Data are mean $\pm$ SD. The listed $p$ value indicates the significance for \model outperforming the best comparison model, with the two-sided $t$-test.
    \textbf{d,} Box plots comparing the AUROC scores of \model, \model-EEG and \model-iEEG on few-shot classification. We perform 3-shot and 8-shot classification for each dataset.
    \textbf{e,} Average improvement of \model over \model-EEG and \model-iEEG on cross-domain evaluation and few-shot classification.
    We first calculate the relative improvement for each experiment and then compute the average of them.
    \textbf{f,} Bar plots comparing the AUROC and BACC scores of \model, \model-EEG and \model-iEEG on out-of-domain recording type evaluation.
    We collect a ECG dataset (Apnea-ECG) with a sleep apnea detection task. Data are mean $\pm$ SD.
    \textbf{a,d,f,} Each experiment is conducted with $n$-fold cross validation ($n$ is the number of subject groups), where we repeat five runs for each fold.
}

\end{center}
\end{figure}
\subsection{Analysis of joint pretraining}
To the best of our knowledge, \model is the first foundational model that combines invasive and non-invasive neural data. 
We next examine the effectiveness of this joint pretraining strategy
to determine whether it is more effective to pretrain a separate model for each recording type and apply it for corresponding downstream tasks with the same recording type, or to utilize a joint pretraining approach.
To this end, we separately pretrained two model variants, namely \model-EEG and \model-iEEG, using EEG and iEEG data, respectively, while keeping the model and pretraining configurations (Supplementary Tables 18 and 19) identical to \model. 
The numbers of patches for pretraining \model-EEG and \model-iEEG are relatively balanced (1.74 billion versus 1.42 billion).
Subsequently, we conducted cross-domain evaluation and few-shot classification on both model variants, in which \model-EEG was evaluated on EEG datasets and \model-iEEG was evaluated on iEEG datasets.

Through a series of experiments, we observed that \model outperformed the other two variants in almost all tasks and experimental settings (Fig.~\ref{fig4}a-d and Extended Data and Figs.~\ref{cross_sub_others} and~\ref{few_shot_others}), except in experiment Absence-16 to Atonic-5 where \model is slightly lower than \model-EEG in terms of BACC (44.55\% versus 47.23\%).
From the overall results, we discovered that the integration of iEEG during joint pretraining enhanced the performance on downstream tasks based on EEG data and vice versa, suggesting that \model is able to capture fundamental insights about brain activities by combining two distinct types of signals.
In comparison to the average improvement over two model variants (Fig.~\ref{fig4}e), the improvement of the \model over \model-EEG was more significant than the improvement over \model-iEEG (7.44\% versus 2.58\% on cross-domain evaluation and 8.13\% versus 6.92\% on few-shot classification in terms of AUROC).
The phenomenon suggested that the boost in performance achieved by incorporating iEEG data in the pretraining was more pronounced, which might be attributed to the lower signal-to-noise ratio and higher accuracy of intracranial neural signals.

Given that joint pretraining leads to increase in performance, we further explored the underlying reasons by analyzing the representations learned by the models.
First, we validated whether joint pretraining can learn more enriched information.
For this purpose, we performed principal component analysis (PCA) to the pretrained representations, selecting principal components until 99\% of the variance could be explained.
We conducted analysis on 15 datasets and recorded the number of principal components $k$ of each dataset.
On 13 out 14 datasets, \model yielded a higher $k$ value (Supplementary Tables 20 and 21), indicating that \model demonstrated the ability to extract more enriched information from downstream datasets.
Furthermore, we conducted an additional experiment by evaluating the performance of the \model versus other variants on another type of biosignal, electrocardiogram (ECG), through which we aimed to verify if joint pretraining results in a stronger adaptability on new tasks due to the acquisition of more general patterns.
The results (Fig.~\ref{fig4}f) showed that \model achieves an improvement of 9.90\% and 19.48\% in terms of AUROC over \model-iEEG and \model-EEG, respectively, demonstrating that joint pretraining enables better generalization to unseen data types. 
To summarize, \model learned richer semantic information and more general patterns of the data than other model variants with only one type of data.
This finding opens up possibilities for expanding signal types and developing more versatile foundational models for biosignals.

\section{Discussion}

We have introduced \model, a brain signal foundation model that learns robust representations of electrical brain recordings for a broad range of clinical applications. 
To the best of our knowledge, \model is the first model pretrained on a large-scale dataset composed of recordings from both invasive and non-invasive modalities, which comprised more than 3 billion signal patches from approximately 16,000 individuals.
We employed a masked modeling strategy to pretrain \model, enabling the model to reconstruct the complete sequence from partial observations.
The architectural design accommodated brain recordings of varying lengths, sampling rates, and electrode counts, thereby enhancing the flexibility of \model for joint pretraining and deployment on EEG and iEEG data.
In comprehensive experiments involving cross-domain evaluation and few-shot classification, we demonstrated the versatility of \model across a wide array of medical and healthcare applications under diverse conditions.
This suggests that \model holds significant potential to support clinical diagnostics and the identification of health conditions.

Beyond its practical value, \model yields valuable insights into the fundamental model governing multimodal neural electrical signals.
The present study represents the initial investigation to utilize joint pretraining with both EEG and iEEG data.
Through rigorous assessment, we demonstrated the effectiveness of this novel methodological approach.
The results of our study demonstrate a significant performance boost derived from the joint pretraining approach, compared to control model configurations. 
Additionally, we have elucidated the factors contributing to the enhanced performance associated with joint training.  This work paves the way for exploring the application of analogous techniques to other medical data domains, where the development of efficient and scalable models is crucial. 
Furthermore, this study establishes a basis for further research into the joint pretraining of models across diverse biological data modalities. 
Future investigations can build upon our findings to examine whether comparable performance enhancements can be attained with other neural data types or even cross-domain data encompassing varied physiological signals.


As an exciting interdisciplinary research direction between neuroscience and artificial intelligence, \model may also provide a new approach to deciphering the mechanisms of brain information processing. 
A key challenge in neuroscience research involves analyzing and elucidating the fundamental operating principles governing population-level neural networks, as derived from high-volume neural recording data. 
Large language models have demonstrated its capacity to extract fundamental attributes of human cognition by compressing substantial human linguistic data~\ref{yin2024entropylawstorydata, naveed2024comprehensiveoverviewlargelanguage}. 
Similarly, the impressive performance of \model, as shown in the present study, may be attributed to their ability to thoroughly comprehend and effectively extract features of brain neural activity. 
Consequently, \model may serve as a digital counterpart to brain biological neural networks. 
Analyzing the network properties of \model could potentially offer insights to guide research on biological neural networks. 
Specifically, by modulating various parameters of \model, analogous models representing diverse brain disorders can be generated, thereby establishing a novel experimental framework for neuroscience and brain disease research.

Despite the promising results, there is a wealth of potential for further development and advancement.
Firstly, \model cannot handle data from other modalities, such as magnetic resonance imaging (MRI), which can provide higher spatial resolution compared to electrical signals and is widely used in healthcare applications.
Consequently, our ultimate objective is to develop a model framework that can accommodate various data modalities.
Secondly, diverse medical scenarios collect various physiological signals, with diagnoses sometimes relying on multiple signal types.
For instance, stroke diagnosis and rehabilitation often require the recording of EEG and EMG (Electromyography)\cite{Jo2022EEG}.
To enhance suitability for a more expansive set of healthcare applications, the model should possess the capability to accommodate a more diverse array of biosignals.  This investigation involved initial efforts utilizing electrocardiogram data, which indicate that additional advancements are necessary before constructing a comprehensive model able to accommodate a variety of physiological measurements.

\section{Methods}


\subsection{Technique details}
Fig.~\ref{fig1}b shows the overall architecture of \model, which is composed of three main components: embedding layer, Transformer encoder and channel attention.
This section aims to introduce the details of these components.

\vpara{Embedding layer. }
The embedding layer of \model projects the original signals into latent embeddings, in which each channel is operated independently.
Given a single-channel brain recording, we first divided the signal into a series of consecutive non-overlapping 1-second patches $\mathbf{P}_i \in \mathbb{R}^{N\times P}$, where $N$ is the number of patches, $P$ is the number of timestamps in each 1-second patch, and $i$ is the channel index.
Then we calculated the time-frequency representations of each patch, in which we chose the spectrograms with Gaussian window.
By keeping the ratio of the window size and the hop size to the patch length $P$ constant, we can align recordings with different sampling rates onto spectrograms with consistent temporal and frequency resolutions.
Specifically, we set the window size equal to $\frac{P}{4}$ and the hop size equal to $\frac{P}{8}$, and the resulting spectrograms are denoted as $\mathbf{S}_i \in \mathbb{R}^{N\times T\times F}$, where $T$ is the length of the time axis and $F$ is the length of the frequency axis.
We convolved $\mathbf{S}_i$ using 2D convolutional kernels to obtain the feature maps $\mathbf{M}_i \in \mathbb{R}^{N\times C_{\text{out}}\times T_{\text{out}} \times F_{\text{out}}}$, where $C_{\text{out}}$ is the output channels and $T_{\text{out}}\times F_{\text{out}}$ is the size of the feature maps. 
Since our design ensured that data with different sampling rates have the same temporal and frequency resolution, the length of the time axis $T_{\text{out}}$ in the resulting feature maps was the same (because the duration of all patches is 1 second), while the length of the frequency axis $F_{\text{out}}$ varied. 
Therefore, we performed padding or truncation on the frequency axis to standardize the size of the feature maps.
Then we flattened the standardized feature maps and projected them with a linear layer to derive the input embeddings $\mathbf{E}_i \in \mathbb{R}^{N\times D}$, where $D$ is the hidden size.

\vpara{Transformer encoder. }
The Transformer encoder is composed of stacked Transformer blocks with bidirectional self-attention, which captures the temporal relationship among the patches within a sequence.
We first concatenated the input embeddings $\mathbf{E}_i$ with a \texttt{[CLS]} token, then added a set of learnable positional embeddings $\mathbf{PE}_i \in \mathbb{R}^{(N+1)\times D}$ to obtain the input of the Transformer encoder.
Like the embedding layer, the Transformer encoder encoded each channel independently and generated a set of outputs $\mathbf{O}_1, \mathbf{O}_2,...,\mathbf{O}_C$, where $C$ is the number of channels.
We derived the whole output $\mathbf{O} \in \mathbb{R}^{C\times (N+1)\times D}$ by concatenating the outputs together.

\vpara{Channel attention. }
The channel attention module aims at capturing the correlation between different channels. 
Specifically, the input $\mathbf{O}^j \in \mathbb{R}^{C\times D}, j=0,1,...,N$ contained $C$ different patches at the same time, which was then performed with a bidirectional self-attention operation.
The output of the channel attention, denoted as $\mathbf{Z} \in \mathbb{R}^{C\times (N+1)\times D}$, served as the latent representation of \model, where $\mathbf{Z}^0$ were sequence-level representations (representations of \texttt{[CLS]} tokens) and $\mathbf{Z}^1,\mathbf{Z}^2,...,\mathbf{Z}^N$ were patch-level representations.

\subsection{Pretraining}
\vpara{Data curation. }
We curated large collections of unannotated electrical brain recordings for pretraining, totaling 13.79 TB data over a duration of 40,907 hours.
The iEEG data were obtained from CCEP~\cite{ccep2023} and a private corpus collected by ourselves, comprising 10.63 TB of data. 
The recordings spanned a duration of 5231 hours and were collected from 91 subjects, ranging in age from 4 to 51 years. 
The sampling rate ranged from 1000 Hz to 4096 Hz, and the number of channels varied from 48 to 238.
The EEG recordings consisted of CAP~\cite{CAP}, HMC~\cite{HMC}, Siena~\cite{Siena}, SRM~\cite{SRM}, TUEG~\cite{TUEG}, Schizophrenia-81, Sleep-EDF~\cite{Sleepedfx}, Stroke-50~\cite{stroke-EEG}, PD-31~\cite{UCSD}, IowaDataset, UNMDataset, AD-184~\cite{FSU}, and a private EEG corpus, with a total of 3.16 TB of data.
The recording duration of the data reached 35,675.5 hours and involved 15,906 subjects, ranging in age from less than 1 year to over 90 years. 
The sampling rate ranged from 100 Hz to 1024 Hz, and the number of channels varied from 1 to 64.

The preprocessing of the pretraining data primarily involved channel selection, downsampling and filtering.
Due to potential equipment issues during the data collection process, there might be invalid channels where no valid brain signals were captured. 
Therefore, we needed to perform channel selection, in which we visualized the recordings and manually selected the valid dchannels.
We only downsampled signals with a sampling rate higher than 1000 Hz until they were below 1000 Hz, with the remaining signals preserved at their original sampling rates.
Then, we applied a bandpass filter in the frequency range of 0.01 Hz to sfreq/3 Hz, where sfreq represents the sampling rate, and applied a notch filter at 50 Hz or 60 Hz to remove powerline noise.

\vpara{Pretraining details. }
The main backbone of \model is the RoBERTa~\cite{liu2019roberta} encoder architecture. The model had a hidden size of 768 and an intermediate size of 2048, with 10 layers and 16 attention heads.
We applied absolute positional encoding with a maximum sequence length of 61 (60 signal patches along with a $\text{[CLS]}$ token).
We pretrained \model on a total of 3,162,233,694 signal patches, including 1,739,447,411 EEG data patches and 1,422,786,283 iEEG data patches.
\model was trained using the AdamW optimizer~\cite{loshchilov2017decoupled}, with $\beta_1=0.9$, $\beta_2=0.95$, $eps=10^{-5}$. 
For the learning rate scheduling, we utilized a linear warmup of 1000 steps to reach a peak learning rate of $1.0 \times 10^{-5}$, followed by a cosine decay of 30,000 steps to decay the final learning rate to 0. 
The total training steps of \model was 16,600.
We employed gradient accumulation during pretraining, where we accumulated gradients for 16 times of forward and backward before performing a parameter update. 
The training process was conducted on 4$\times\text{A100}$ GPUs with a global batch size of 2,560,000 (patches) and the entire process took 100 hours.

\subsection{Downstream evaluation}
\vpara{Competing methods. }
We compared \model to 3 publicly available models: \labram~\cite{jiang2024large}, \brainbert~\cite{wang2023brainbert} and \moment~\cite{goswami2024moment}.
\labram is an open-weight model pretrained on more than 2500 hours of EEG data. It tokenized the EEG into discrete tokens by training a neural tokenizer, and was pretrained with symmetric masked modeling.
\brainbert is a reusable, off-the-shelf, subject-agnostic, and electrode-agnostic model that provides embeddings for intracranial recordings. It was pretrained on 43.7 hours of iEEG data recorded from 10 subjects.
During pretraining, it masked multiple continuous bands of random frequencies and time intervals in the time-frequency representations.
\moment is a family of open-source foundation models for general-purpose time series analysis.
It was pretrained on a large collection of publicly available datasets from 13 different domains, which included 20.085 GB ($\approx 0.02$ TB) worth of 13 million unique time series and 1.23 billion timestamps (0.15 billion patches).
\moment also adopted a masked modeling strategy by masking and reconstructing the original time series.

\vpara{Evaluation datasets. }
The evaluation benchmark comprised 15 distinct datasets, which contain 7 different healthcare scenarios.

Alzheimer's disease: AD-65~\cite{AD-65} contains the EEG resting state-closed eyes recordings from 88 subjects in total (44 males, ages 53–79; and 44 females, ages 44–79).
For the participants, 36 were diagnosed with Alzheimer's disease (AD group), 23 were diagnosed with Frontotemporal Dementia (FTD group) and 29 were healthy subjects (CN group).
We randomly split the subjects from AD group and CN group into 5 groups.
The data comprised 19 channels with a sampling rate of 250 Hz. 
After processing, we obtained a total of 5349 samples and each sample contains a 10-second data segment. 

Epilepsy: The CHB-MIT~\cite{CHB-MIT, shoeb2009application} database consists of EEG recordings from 22 pediatric subjects (5 males, ages 3–22; and 17 females, ages 1.5–19) with intractable seizures.
We randomly split the subjects into 5 groups.
The data comprised 23 channels with a sampling rate of 256 Hz. 
We split the data into 10-second segments and obtained 4148 samples in total.
Absence-16, Clonic-6 and Atonic-5 are 3 private datasets that were collected from patients with absence seizures, clonic seizures and atonic seizures, respectively.
The annotations were divided into three categories: epileptic waveforms, normal waveforms, and Interictal epileptiform discharge (IED).
The recordings comprised 19 channels with a sampling rate of 256 Hz.
After processing the data into 4-second segments, we obtained a total of 8016 samples from Absence-16, 2426 samples from Clonic-6, and 1587 samples from Atonic-5.
We randomly divided the 16 patients in Absence-16 into 5 groups.
The Mayo-Clinic~\cite{nejedly2020multicenter} data were collected between 1 AM and 3 AM from 25 patients with drug resistant epilepsy (DRE) undergoing evaluation for epilepsy surgery.
The FNUSA~\cite{nejedly2020multicenter} dataset is made up of iEEG data collected in awake resting state from 14 patients diagnosed with DRE.
These patients underwent a standard pre-surgical monitoring for localization of seizure onset zone (SOZ), a standard procedure for epilepsy surgery.
We split Mayo-Clinic and FNUSA into 6 and 5 subject groups, respectively. 
Both Mayo-Clinic and FNUSA were segmented into 3-second data clips and downsampled to 1000 Hz. 
In order to locate the SOZ, annotations were made for each channel.
We preserved the data segments annotated with physiological activity, pathological (epileptic) activity and artifacts.
In total, Mayo-Clinic contained 113,260 samples and FNUSA contained 179,629 samples.
DRE-Clinical is a private dataset that contains the iEEG recordings from 8 patients with DRE. The annotation was also channel-wise for SOZ localization.
We split the patients into 4 groups, with two individuals in each group.
The recordings were segmented into 20-second data clips and we obtained 229,932 samples in total.

Depression: MDD-64~\cite{mumtaz2016mdd} contains EEG recordings from 64 subjects with 34 of them diagnosed with Major Depressive Disorder (MDD). We randomly split the subjects into 5 groups.
The data comprised 19 channels with a sampling rate of 256 Hz. 
We split the data into 10-second segments and obtained 7309 samples in total.
Depression-122~\cite{depression122} consists of resting EEG data with 122 college-age participants (47 males, ages 18–24; 74 females, ages 18–23; and 1 unknown) with their scores in Beck Depression Inventory (BDI). 
According to ~\cite{chang2023depression}, participants with BDI scores $>13$ were considered depressed. Healthy controls had stable low BDI scores ($<7$) and no self-reported history or symptoms of anxiety disorder.
The data comprised 64 channels with a sampling rate of 500 Hz. 
We split the data into 10-second segments and obtained 5836 samples in total.

Schizophrenia: Schizophrenia-28~\cite{Schizophrenia-28} comprises 14 patients with paranoid schizophrenia and 14 healthy controls. Data were acquired with the sampling frequency of 250 Hz using the standard 10-20 EEG montage with 19 EEG channels.
We randomly split the subjects into 5 groups.
For the EEG recordings, we split the data into 10-second segments and obtained 5744 samples.

Attention deficit hyperactivity disorder (ADHD):
ADHD-Adult~\cite{ADHD-adult} was collected from 79 participants, including 42 healthy adults and 37 adults with ADHD (age 20-68 years; male/female: 56/23).
The dataset contained 256 Hz EEG signals recorded from five channels, including O1, F3, F4, Cz, and Fz, with each subject recorded with two channels.
The subjects were randomly split into 5 groups.
We split the data into 5-second segments and obtained 5056 samples in total.
ADHD-Child~\cite{ADHD-child} contains EEG data collected from 121 children (ages 7-12), including 61 with ADHD and 60 healthy controls. 
The EEG recordings were performed based on 10-20 standard by 19 channels at 128 Hz sampling frequency.
We randomly split the subjects into 5 groups.
After processing, we obtained a total of 3322 samples and each sample contains a 5-second data segment.

Sleep Deprivation:
SD-71~\cite{SD-71} provides resting-state EEG data from 71 participants who underwent two experiments involving normal sleep and sleep deprivation.
We randomly split the subjects into 5 groups.
The data comprised 61 channels with a sampling rate of 500 Hz. 
We split the data into 5-second segments and obtained 13,700 samples in total.

Sleep Apnea: Apnea-ECG~\cite{Apnea-ECG} is an annotated database with 70 nighttime ECG recordings. 
Each recording included a continuous digitized ECG signal with a sampling rate of 100 Hz.
We divided the recordings into 5 groups, split the data into 60-second segments, and obtained 34,271 samples in total.

\vpara{Cross-subject evaluation. }
The cross-subject evaluation involved experiments on 12 datasets: 
AD-65 (Alzheimer's disease diagnosis), CHB-MIT (seizure detection), Absence-16 (seizure detection), Mayo-Clinic (seizure detection and SOZ localization), FNUSA (seizure detection and SOZ localization), DRE-Clinical (seizure detection and SOZ localization), MDD-64 (MDD diagnosis), Depression-122 (depression diagnosis), Schizophrenia-28 (schizophrenia diagnosis), ADHD-Adult (ADHD diagnosis), ADHD-child (ADHD diagnosis) and SD-71 (sleep deprivation detection).
We used the AdamW optimizer for fine-tuning, with $\beta_1=0.9$, $\beta_2=0.95$, $eps=10^{-5}$. 
For all the models, we fine-tuned the pretrained encoder and the classification head with a learning rate of $1\times 10^{-5}$ and $1\times 10^{-4}$.
The models were trained for up to 30 epochs, and then the best-performing models on the validation set were selected for testing.


\vpara{Cross-hospital and cross-subtype evaluation. }
The cross-hospital evaluation involved Mayo-Clinic and FNUSA, and the cross-subtype evaluation involved Absence-16, Clonic-6, and Atonic-5.
In the experiments, we fine-tune the models on the source dataset with a fixed number of epochs and directly evaluated on the target dataset.
We used the AdamW optimizer, with $\beta_1=0.9$, $\beta_2=0.95$, $eps=10^{-5}$.
We fine-tuned the pretrained encoder and the classification head for 5 epochs with a learning rate of $1\times 10^{-5}$ and $1\times 10^{-4}$.

\vpara{Few-shot classification. }
The datasets used for few-shot classification were identical to those used for cross-subject evaluation. 
We classified the queries by comparing with prototypes.
Specifically, in $K$-shot, $M$-class classification, given the representations of the support set $\{\mathbf{u}_i^j | i=1,2,...,K; j=1,2,...,M \}$, we obtained the prototypes for each class as the mean of the examples: $\{\mathbf{v}^j=\frac{1}{K}\sum_{i=1}^K\mathbf{u}_i^j | j=1,2,...,M \}$.
Given the representation of a query $\mathbf{z} \in \mathbb{R}^{C\times D}$, where $C$ is the number of channels and $D$ is the hidden size, we calculated the channel-wise cosine similarities between the query representation and prototypes: $\{\text{sim}^j=\cos (\mathbf{z}, \mathbf{v}^j) \in \mathbb{R}^C | j=1,2,...,M\}$.
The scores of the query were the mean of channel-wise similarities and we chose the class with the highest score as the prediction: $y_{\text{pred}}=\text{argmax}([s^1, s^2,...,s^M])$, where $s^j=\sum_{c=1}^C\text{sim}^j_c$.

\subsection{Data Availability}
This study utilized the following publicly available datasets for downstream benchmarking:
AD-65 (\url{https://openneuro.org/datasets/ds004504/versions/1.0.2}),
CHB-MIT (\url{https://physionet.org/content/chbmit/1.0.0/}),
Mayo-Clinic (\url{https://springernature.figshare.com/collections/Multicenter_intracranial_EEG_dataset_for_classification_of_graphoelements_and_artifactual_signals/4681208}),
FNUSA (\url{https://springernature.figshare.com/collections/Multicenter_intracranial_EEG_dataset_for_classification_of_graphoelements_and_artifactual_signals/4681208}),
MDD-64 (\url{https://figshare.com/articles/dataset/EEG_Data_New/4244171}),
Depression-122 (\url{https://openneuro.org/datasets/ds003478/versions/1.1.0},
Schizophrenia-28 (\url{https://repod.icm.edu.pl/dataset.xhtml?persistentId=doi:10.18150/repod.0107441}),
ADHD-Adult (\url{https://data.mendeley.com/datasets/6k4g25fhzg/1}),
ADHD-Child (\url{https://ieee-dataport.org/open-access/eeg-data-adhd-control-children}),
SD-71 (\url{https://openneuro.org/datasets/ds004902/versions/1.0.5},
Apnea-ECG (\url{https://physionet.org/content/apnea-ecg/1.0.0/}).

\subsection{Code Availability}
We will release the model weights, pretraining code, and usage code upon publication.




\section*{Ethics Declarations}
The authors declare no competing interests.

\begin{appendices}
\clearpage
\section{Extended Data}
\begin{figure}[h]
\begin{center}

\includegraphics[width=\linewidth]{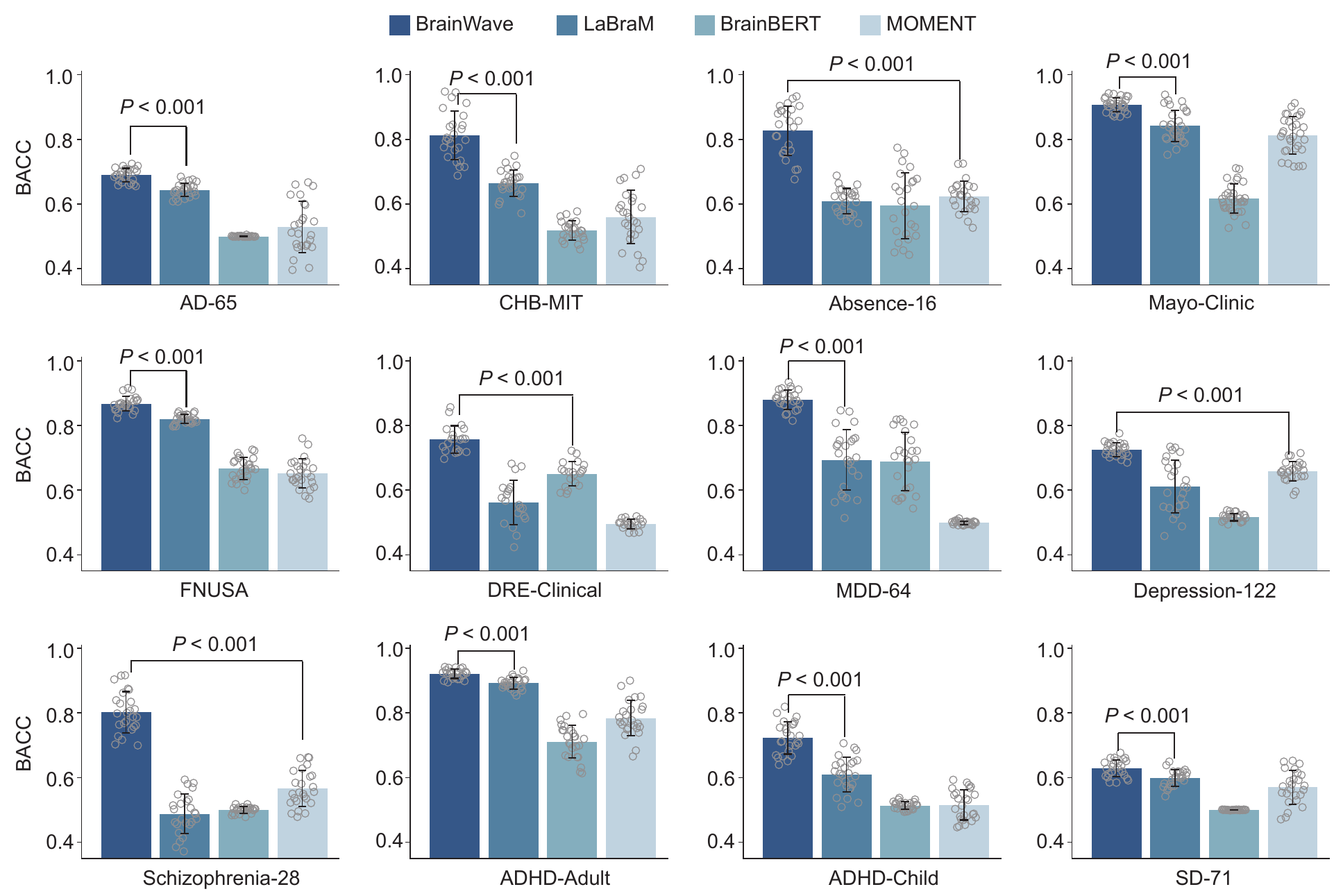}

\caption{
    \label{cross_sub_bacc}
    \textbf{Performance of cross-subject tasks. }
    Bar plots comparing the BACC scores of \model and competing models on cross-subject tasks.
    Data are mean $\pm$ SD. Each experiment is conducted with $n$-fold cross validation ($n$ is the number of subject groups), where we repeat five runs for each fold. The listed $p$ value indicates the significance for \model outperforming the best comparison model, with the two-sided $t$-test.
}

\end{center}
\end{figure}
\clearpage
\begin{figure}
\begin{center}

\includegraphics[width=\linewidth]{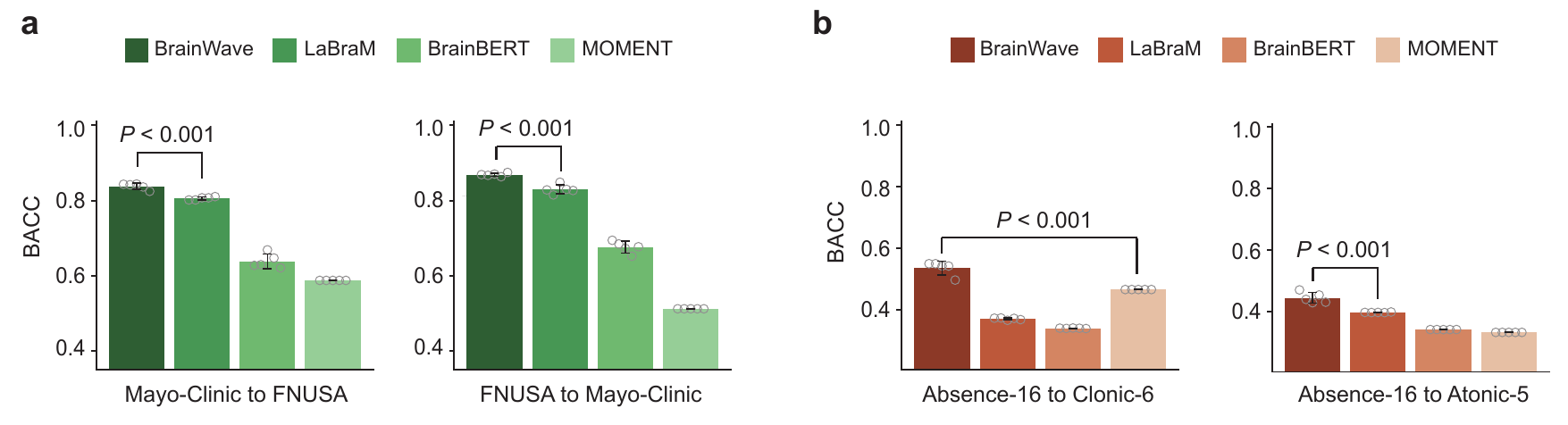}

\caption{
    \label{cross_bacc}
    \textbf{Performance of cross-hospital and cross-subtype tasks. }
    \textbf{a,} Bar plots comparing the BACC scores of \model and competing models on cross-hospital tasks.
    \textbf{b,} Bar plots comparing the BACC scores of \model and competing models on cross-subtype tasks.
    Data are mean $\pm$ SD. Each experiment is repeated five runs. The listed $p$ value indicates the significance for \model outperforming the best comparison model, with the two-sided $t$-test.
}

\end{center}
\end{figure}
\clearpage
\begin{figure}
\begin{center}

\includegraphics[width=\linewidth]{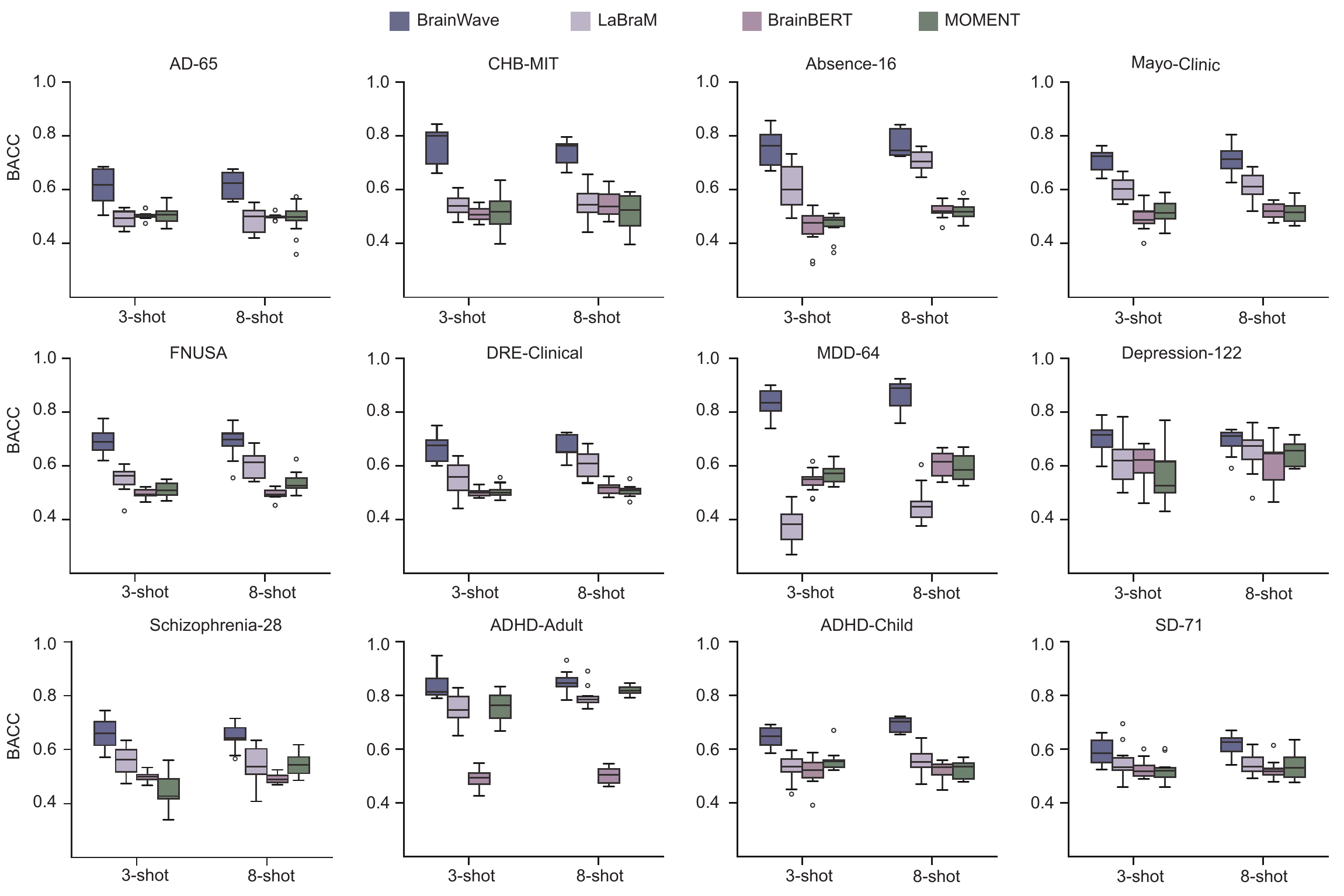}

\caption{
    \label{few_shot_bacc}
    \textbf{Performance of few-shot classification. }
    Box plots comparing the BACC scores of \model and competing models on few-shot classification. We conduct $n$-fold cross validation for each experiment and repeat five runs per fold. We perform 3-shot and 8-shot classification for each task.
}

\end{center}
\end{figure}
\clearpage
\begin{figure}
\begin{center}

\includegraphics[width=\linewidth]{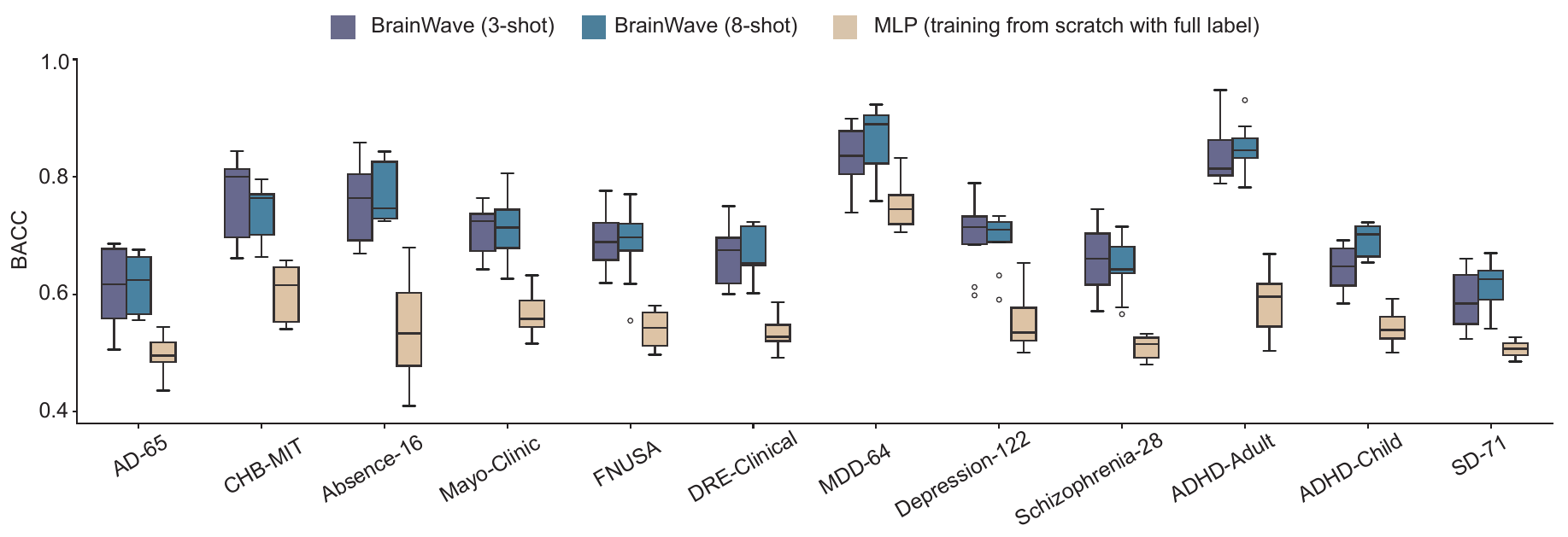}

\caption{
    \label{mlp_bacc}
    \textbf{Comparison between few-shot classification with \model and end-to-end MLP.}
    Box plots comparing the BACC scores of \model on few-shot classification and end-to-end trained MLP with full-label supervision.
}

\end{center}
\end{figure}
\clearpage
\begin{figure}
\begin{center}

\includegraphics[width=\linewidth]{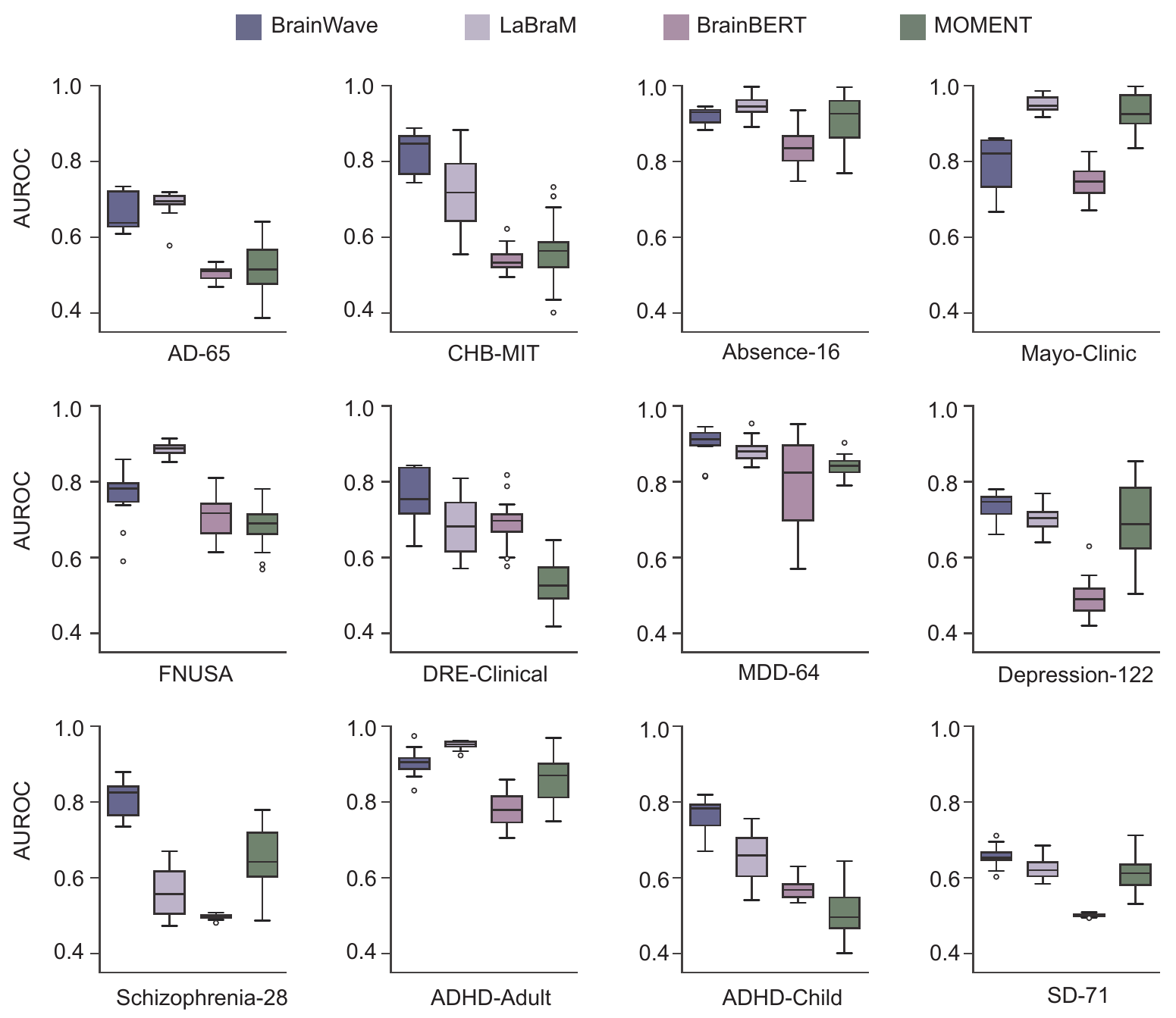}

\caption{
    \label{8shot_finetune_auc}
    \textbf{Comparison between few-shot classification with \model and full-label fine-tuning of competing models.}
    Box plots comparing the AUROC scores of \model on 8-shot classification and other pretrained models on full-label fine-tuning. 
    For all the models, we conduct $n$-fold cross validation in each experiment and repeat five runs per fold.
}

\end{center}
\end{figure}
\clearpage
\begin{figure}
\begin{center}

\includegraphics[width=\linewidth]{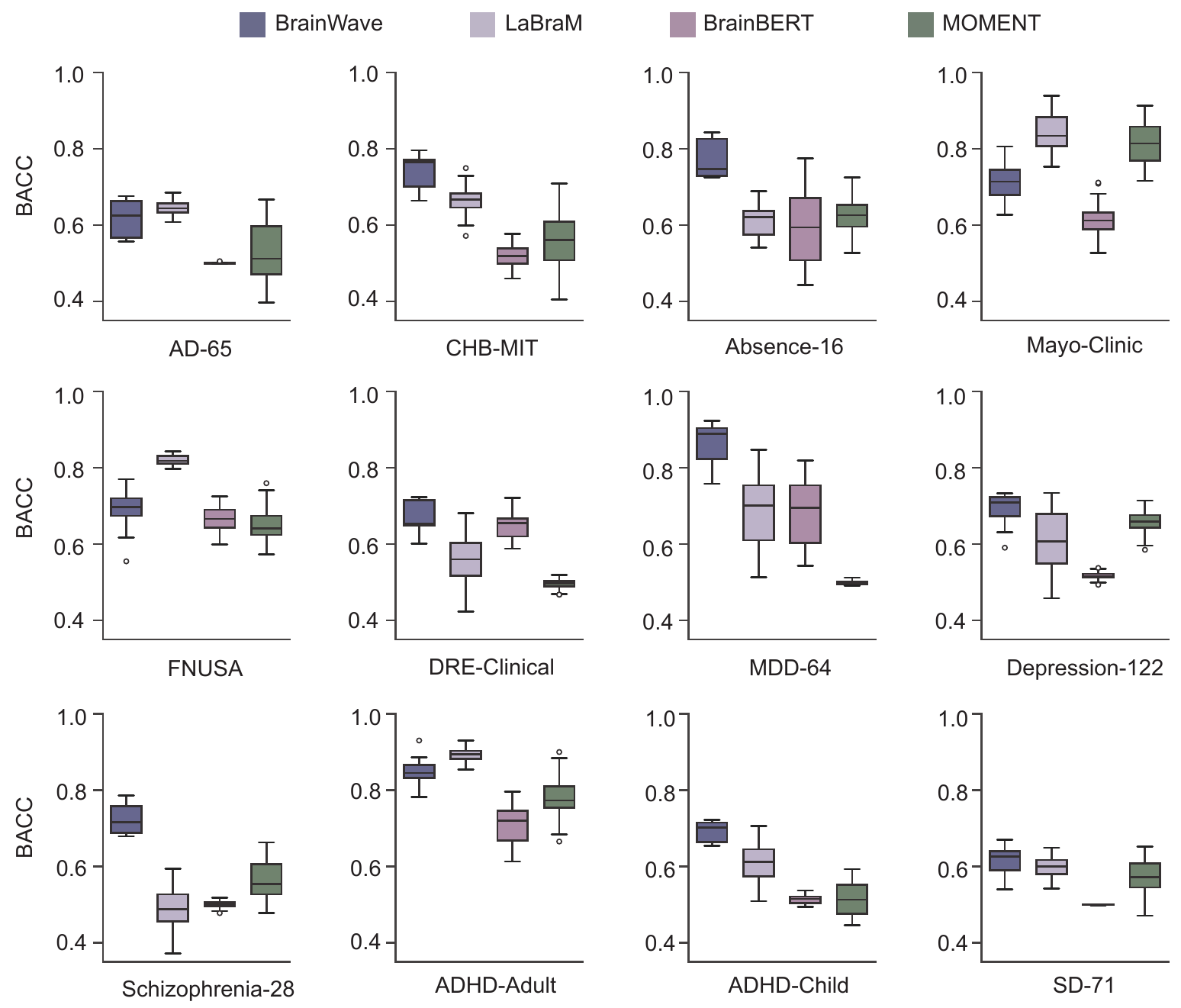}

\caption{
    \label{8shot_finetune_bacc}
    \textbf{Comparison between few-shot classification with \model and full-label fine-tuning of competing models.}
    Box plots comparing the AUROC scores of \model on 8-shot classification and other pretrained models on full-label fine-tuning.
    For all the models, we conduct $n$-fold cross validation in each experiment and repeat five runs per fold.
}

\end{center}
\end{figure}
\clearpage
\begin{figure}
\begin{center}

\includegraphics[width=\linewidth]{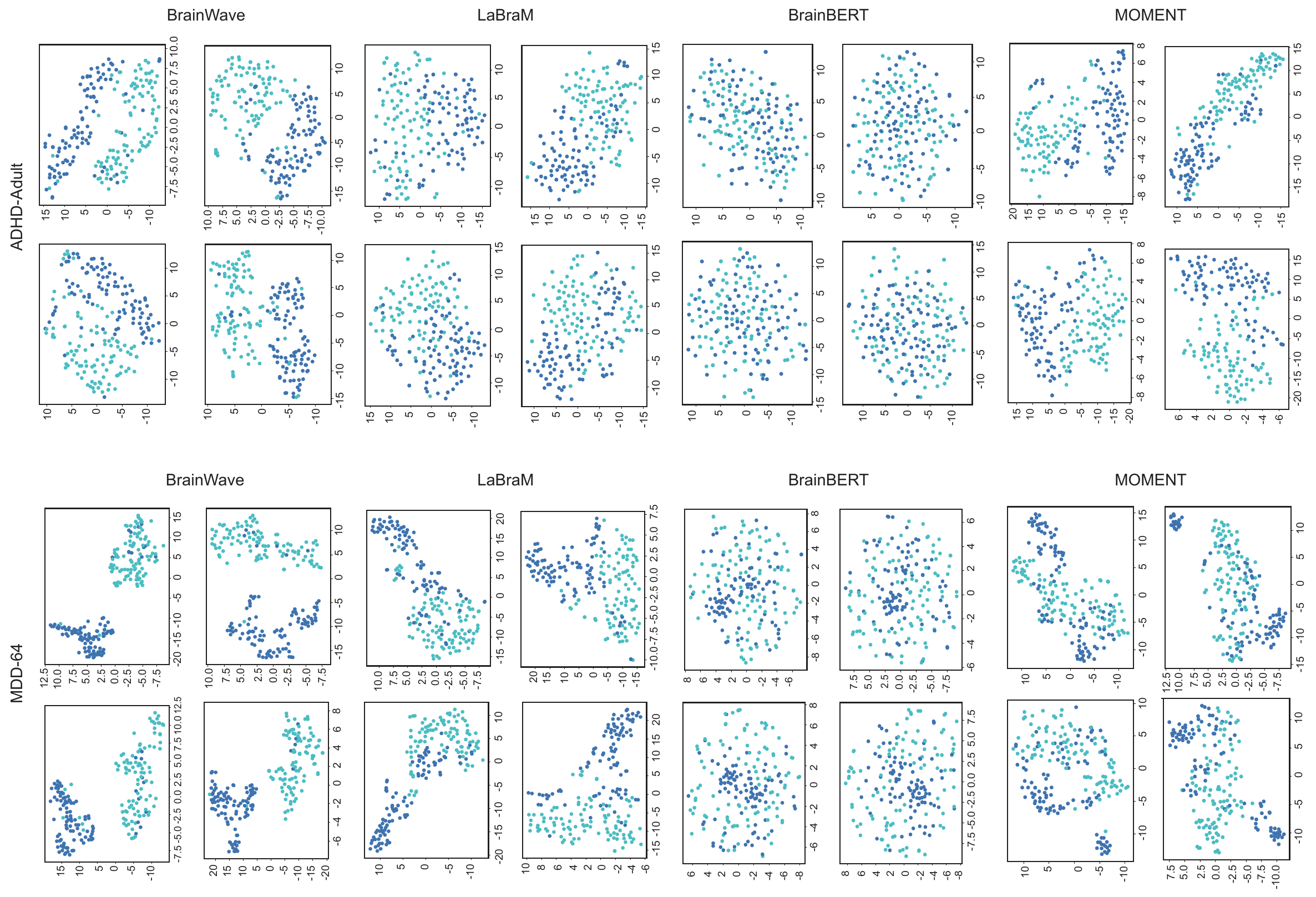}

\caption{
    \label{tsne}
    \textbf{t-SNE analysis of few-shot classification. }
    t-SNE plots of the pretrained representations on ADHD-Adult and MDD-64 generated from \model and other pretrained encoders. Each model contains four subplots, with each subplot generated by randomly sampling a portion of the original dataset.
}

\end{center}
\end{figure}
\clearpage
\begin{figure}
\begin{center}

\includegraphics[width=\linewidth]{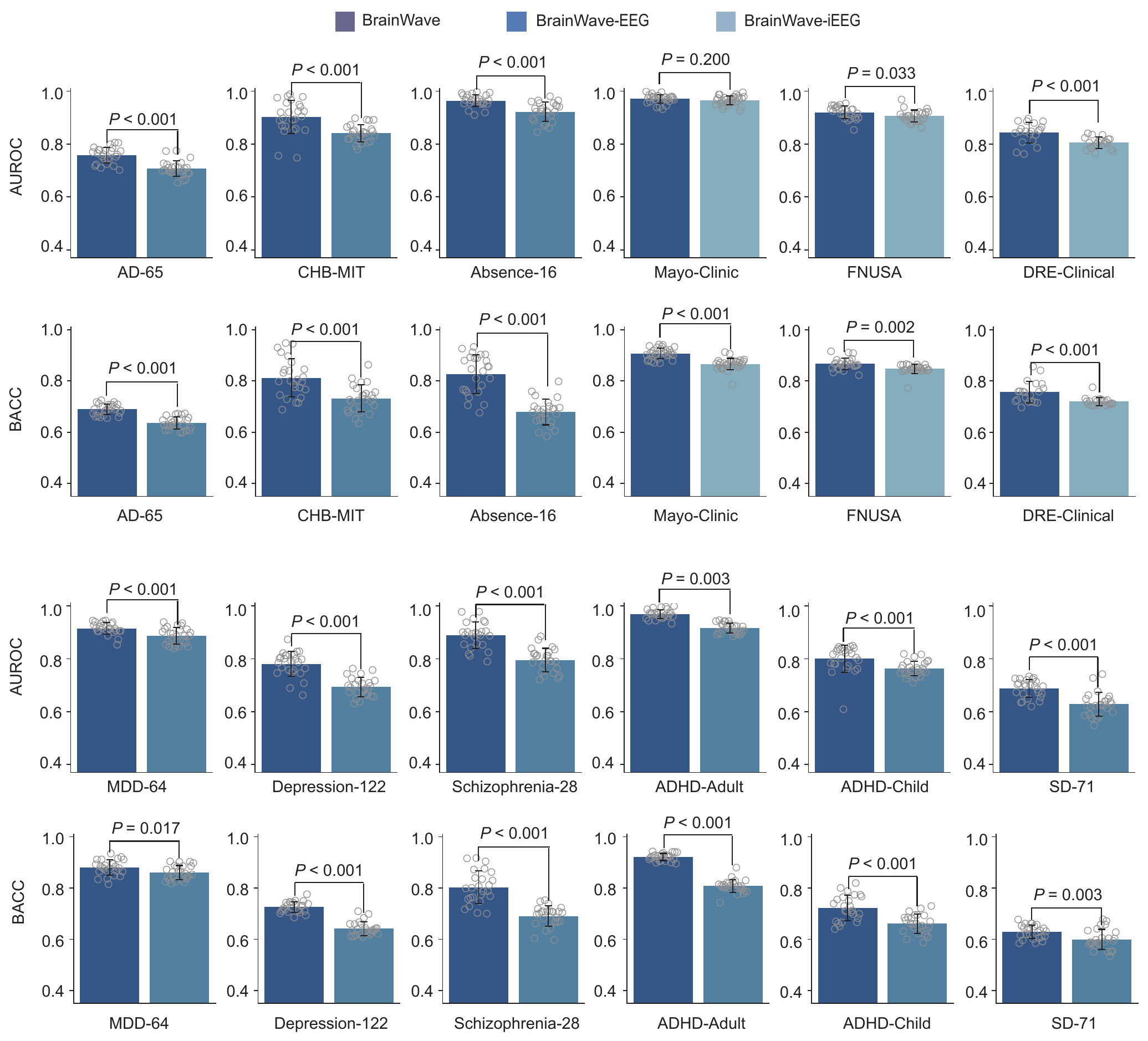}

\caption{
    \label{cross_sub_others}
    \textbf{Performance of cross-subject evaluation with \model, \model-EEG and \model-iEEG.}
    Bar plots comparing the AUROC and BACC scores of \model, \model-EEG and \model-iEEG on cross-subject tasks.
    Each experiment is conducted with $n$-fold cross validation ($n$ is the number of subject groups), where we repeat five runs for each fold. 
}

\end{center}
\end{figure}
\clearpage
\begin{figure}
\begin{center}

\includegraphics[width=\linewidth]{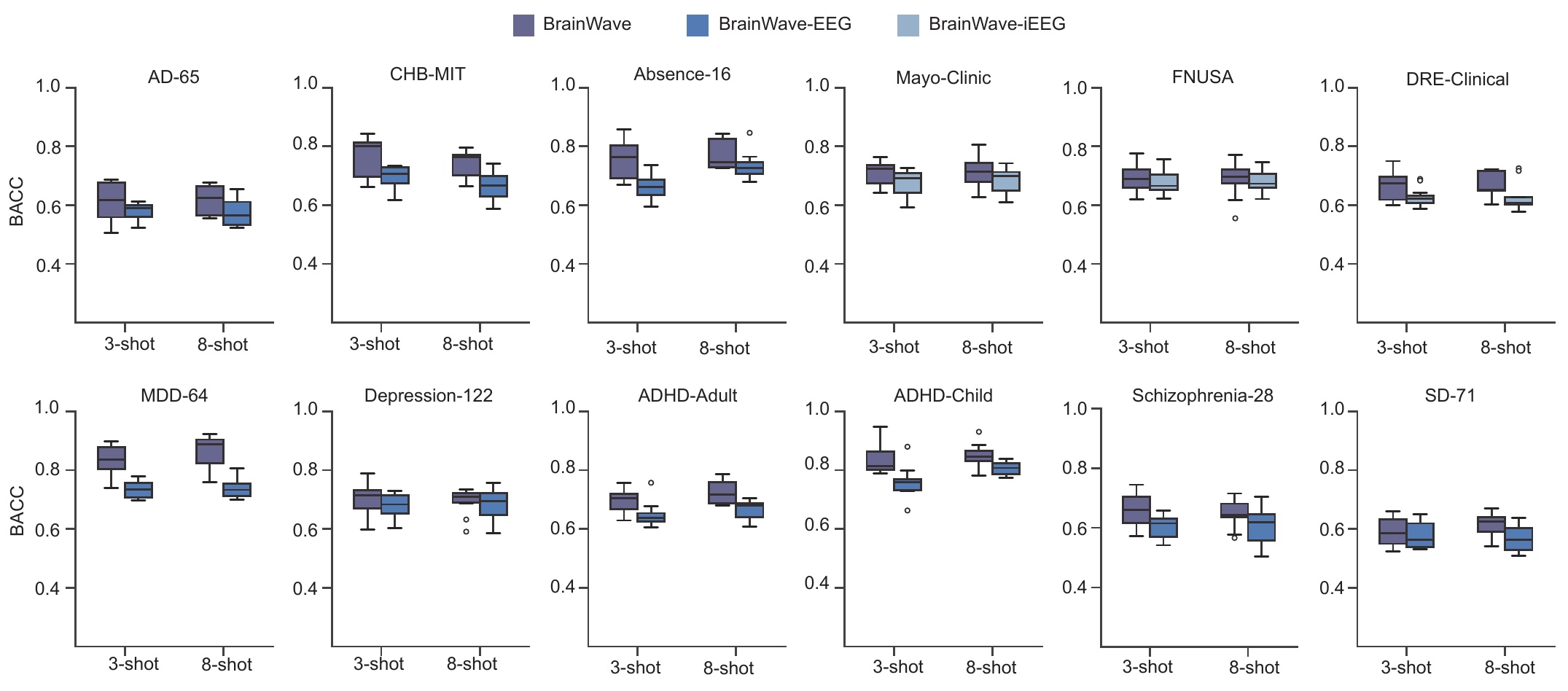}

\caption{
    \label{few_shot_others}
    \textbf{Performance of few-shot classification with \model, \model-EEG and \model-iEEG. }
    Box plots comparing the BACC scores of \model, \model-EEG and \model-iEEG on few-shot classification. We perform 3-shot and 8-shot classification for each task. Data are mean $\pm$ SD. The listed $p$ value indicates the significance for \model outperforming the best comparison model, with the two-sided $t$-test.
}

\end{center}
\end{figure}
\clearpage
\begin{figure}
\begin{center}

\includegraphics[width=\linewidth]{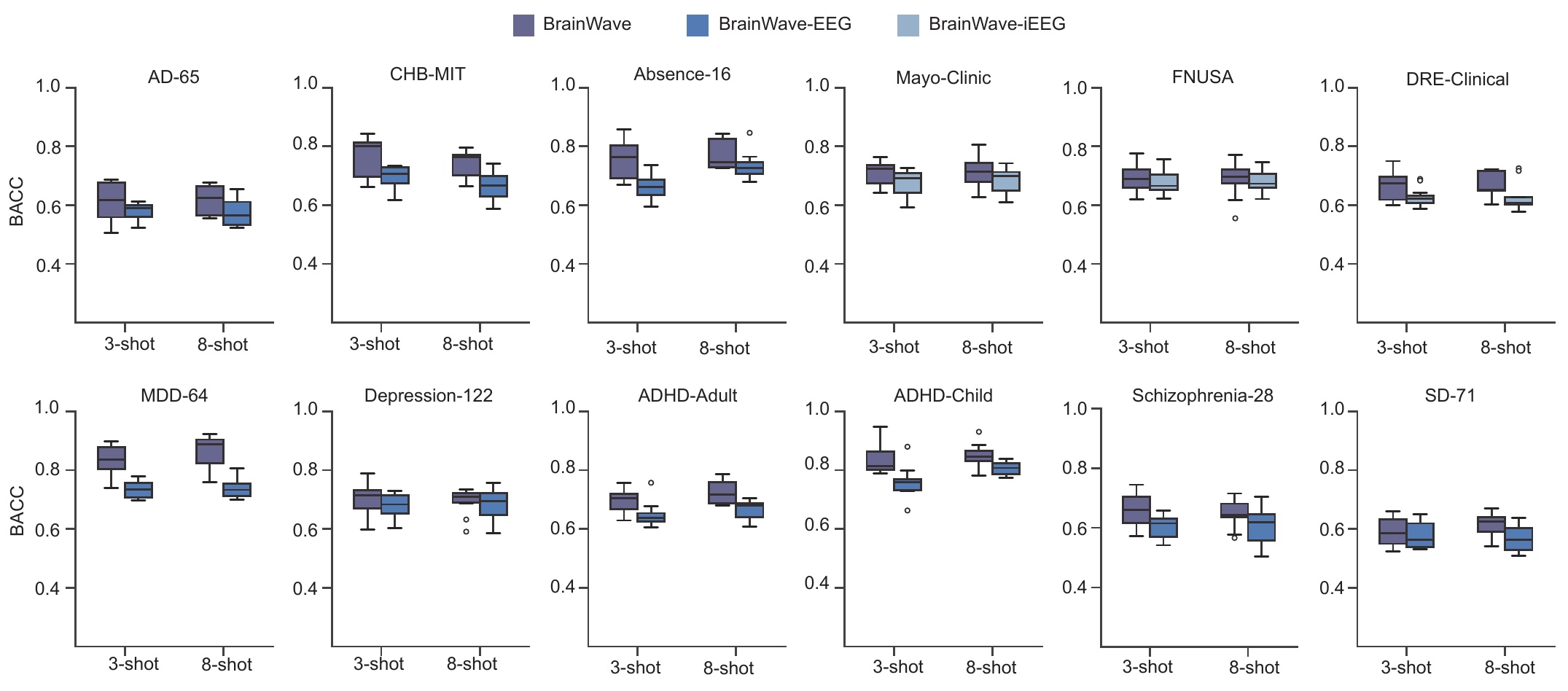}

\caption{
    \label{few_shot_others}
    \textbf{Performance of few-shot classification with \model, \model-EEG and \model-iEEG. }
    Box plots comparing the BACC scores of \model, \model-EEG and \model-iEEG on few-shot classification. We perform 3-shot and 8-shot classification for each task. Data are mean $\pm$ SD. The listed $p$ value indicates the significance for \model outperforming the best comparison model, with the two-sided $t$-test.
}

\end{center}
\end{figure}
\end{appendices}

\end{document}